# Recovery Risk: Application of the Latent Competing Risks Model to Non-performing Loans

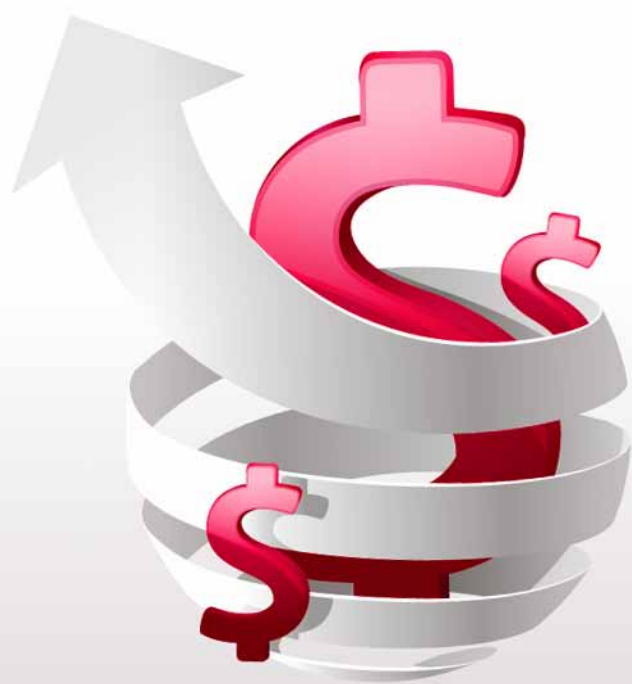

**Mauro R. Oliveira**
**Francisco Louzada**

## Abstract

This article proposes a method for measuring the latent risks involved in the recovery process of non-performing loans in financial institutions and/or business firms that deal with collection and recovery processes. To that end, we apply the competing risks model referred to in the literature as the promotion time model. The result achieved is the probability of credit recovery for a portfolio segmented into groups based on the information available. Within the context of competing risks, application of the technique yielded an estimation of the number of latent events that concur to the credit recovery event. With these results in hand, we were able to compare groups of defaulters in terms of risk or susceptibility to the recovery event during the collection process, and thereby determine where collection actions are most efficient. We specify the Poisson distribution for the number of latent causes leading to recovery, and the Weibull distribution for the time up to recovery. To estimate the model's parameters, we use the maximum likelihood method. Finally, the model was applied to a sample of defaulted loans from a financial institution.

**Keywords:** Competing Risks, Credit Recovery, Default.

## 1. Introduction

Statistical methods are used at almost every stage of a successful business. In the financial industry, initially, market surveys are applied for new product launches, followed by Scorecard models to grant credit to new customers, Behavior Scoring models to increase loyalty and revenues per customer, and finally Collection Scoring models, which are statistical techniques intended to optimize the process of collecting and recovering credits in default.

Therefore, statistical models enable automating process, which is crucial for the industry's players to maintain a high portfolio growth.

In this paper, we propose to contribute to the employment of a statistical method that, as far as the authors are aware, has not been applied to analysis of the collection process at a financial institution. The results obtained enable confirming how and when collection actions are most efficient for the bank and, therefore, adding inputs to propose recovery process improvements.

For this particular effort, we have available a dataset made up of approximately 22 thousand loans from a financial institution that entered default between 2009 and 2011. The rule to characterize default was the same for all customers, that is, 90 days past due payment on the loans installments. The same collection process was applied to every contract, that is, they were all subjected to the same actions on the part of the collections department. The financial institution that provided the data kept secrecy over the collection methods used.

The collection process chosen by the institution considered a 24-month workout period for defaulted loans and, based on a collection rule, certain steps were taken in an attempt to recover non-performing credits.

To apply the proposed methodology, we will only consider fully recovered contracts along with totally lost contracts. That is, if a contract has been partly recovered at the end of the 24-month period, it will not enter the database for application of the method.

Thereafter, our database includes information on time up to full recovery of the contract, and on the other hand, information on fully lost contract. In this latter case, obviously, the observations on date of recovery were not gathered. According to survival analysis terminology, those times were regarded as censored.

Table 1 summarizes the total number of contracts that make up the database available for modeling. They include 22,109 defaulted contracts, of which approximately 64% had not been recovered by the end of the 24-month recovery period.

The bank only made available two items of customer information, or model co-variables. One concerns the customer's risk profile, referred to as "Behavior Score range" (FX-BS), which returns the values 1, 2, 3 and 4; the other one has to do with the contracted product and is called "contract amount range" (FX-CV), also returning the values 1, 2, 3 and 4.

To more clearly illustrate the results obtained and enable easier comparison of inter-group susceptibility to recovery, we only consider some ranges of the co-variables available. The data on tables 1, 2 and 3 (developed by the authors) indicate that customer profiles with Behavior Score range equal to 2 and contract amount range equal to 2 show the highest rates of recovery. These results find support in those obtained in Section 3, with the application of the competing risks model.

■ Table 1

| Group | Recovered | Unrecovered | % Non-recovery | Average recovery time (months) |
|---|---|---|---|---|
| Population: 22,109 | 8.047 | 14.062 | 63,60% | 9,85 |
| Range 1 – Contracted amount: 5,532 | 2.036 | 3.496 | 63,19% | 10,35 |
| Range 2 – Contracted amount: 5,478 | 2.552 | 2.926 | 53,41% | 18,94 |
| Range 1 – Behavior Score: 7,245 | 1.719 | 5.526 | 76,27% | 11,69 |
| Range 2 – Behavior Score: 5,503 | 3.280 | 2.223 | 40,39% | 21,63 |

■ Table 2

| Sub-group | Recovered | Unrecovered | % Non-recovery | Average recovery time (months) |
|---|---|---|---|---|
| Subpopulation: Range 1 – Contracted amount: 2,895 | 1.203 | 1.692 | 58,44% | 10,84 |
| Range 1 – Behavior Score: 1,338 | 347 | 991 | 74,06% | 11,99 |
| Range 2 – Behavior Score: 1,557 | 856 | 701 | 45,02% | 20,16 |

■ Table 3

| Sub-group | Recovered | Unrecovered | % Non-recovery | Average recovery time (months) |
|---|---|---|---|---|
| Subpopulation: Range 2 – Contracted amount: 3,270 | 1.694 | 1.576 | 48,19% | 9,19 |
| Range 1 – Behavior Score: 1,827 | 618 | 1.209 | 66,17% | 10,66 |
| Range 2 – Behavior Score: 1,443 | 1.076 | 367 | 25,43% | 19,38 |

Based on this data structure, that is, with information on the occurrence or non-occurrence of an event and the time up to this occurrence, one may apply the statistical methodology known as survival analysis. Development of the theory and its application to real data are widely discussed in the literature, particularly in the medical area, where, for example, studies have been conducted on the survival period of patients subjected to different kinds of treatments and drugs. We recommend Maller & Zhou (1996) and Ibrahim et al. (2001) to interested readers.

Competing risks modeling, which is this article's purpose, is widely known in the literature and has been extensively discussed in papers such as Cooner et al. (2006), Cooner et al. (2007), Xu et al. (2011). In addition to the large number of additional papers, this modeling has been gaining importance due, mainly, to the work of Chen et al. (1999), Tsodikov et al. (2003), and Tournoud & Ecochard (2007).

As far as the authors' knowledge, competing risks modeling has not yet been applied to modeling the risk behavior of credit portfolios during the recovery process. Therefore, by analyzing the risk behavior leading to the event of interest – recovery – we may then calculate the probability of recovery for a given contract within the chosen period.

With the results of the modeling in hand, we compare the estimated latent risks in the process leading to recovery for different customer groups and the respective probability of recovery. Our main objective was to identify the characteristics of customers that result in greater efficiency in the recovery process.

The following sections are organized as follows:

    2 – competing risks model and how the model's parameters are estimated;
    3 – application of the model to the database;
    4 – conclusions and discussion of the results.

## 2. Model Formulation

In the formulation of a competing risks model, recovery, or any other relevant event, is regarded as a result brought about by causes that operate concurrently over time. Therefore, two statistical distributions are attributed to formulate this model: one for the random variable "time up to event" and another for a random variable that models the number of competing events. For an in-depth study of the matter, we recommend the books of Crowder (2010) and Pintilie (2006), among others.

Our dataset is made up of a population of approximately 22 thousand defaulted contracts from a Brazilian financial institution's credit portfolio. All contracts are subject to the same collection rule, that is, the same collection actions were implemented over a recovery process at most 24 months in length. The recovery process at hand resulted in one of two situations: time to full recovery of the non-performing loans and, for unrecovered ones, as in survival analysis, time is considered to be censored at month 24.

In this paper, we use the probability distributions most frequently employed in survival analysis and competing risks modeling literature. We assume that the time to the event follows the Weibull distribution, represented by the random variable T, and that the number of risk events follows the Poisson distribution, represented by the random variable **M**.

The next formulation assumes that some clients may not be susceptible to recovery, so that the number of competing risks for the recovery event may be zero. The model is generally known as the promotion time model, and has appeared in the literature in previous works like Chen et al. (1999) and Yakovlev & Tsodikov (1996).

Therefore, we assume **M** is Poisson distributed with a probability mass function given by:

$$P(M=m) = \frac{\theta^m \exp(-\theta)}{m!}$$

where $\theta > 0$ e $m = 0,1,2,\ldots$.

For every **i=0, 1, 2, ..., m**, let $T_i$ be the random variable due to the ith risk factor leading to recovery, which is also assumed to be independent from the number of risks given by **M**. The variable $T_i$ is assumed to folllow the Weibull distribution, whose probability function is given by:

$$f(t) = \gamma \beta^\gamma t^{\gamma-1} \exp(-(\beta t)^\gamma)$$

where $t > 0$, $\gamma > 0$ e $\beta > 0$.

Therefore, the time of the occurrence of the event is defined as the minimum time out of all m risk factors, that is, $Y = \min(T_0, T_1, \ldots, T_m)$.

As shown in Bereta et al. (2011), Chen et al. (1999) and Yakovlev & Tsodikov (1996), the random variable **Y**, probability density function is:

$$f_Y(t) = \theta f(t) \exp[-\theta(F(t))]$$

In the same reference, the authors give the survival function as

$$S_Y(t) = \exp[-\theta(F(t))]$$

As expected, the database has a large number of unrecovered contracts. The literature regards these as immune to the event and, therefore, in our case, they are regarded as contracts lost due to default. The output of the model that provides an estimate for this value is known as cure fraction, and is given by **exp(-θ)** in this case.

To estimate the model's parameters, we use the maximum likelihood estimation with the presence of censored events. The censure indicator is such that $\delta_i = 1$ if the contract is recovered and $\delta_i = 0$ otherwise. Therefore, the likelihood function is given as:

$$L(\Theta \mid t) = \prod_1^n f_Y(\Theta \mid t)^{\delta_i} S_Y(\Theta \mid t)^{1-\delta_i}$$

onde $\Theta = (\theta, \gamma, \beta)$.

## 3. Application

Competing risks modeling enables lenders to have practical interpretation of the parameters obtained. The Poisson parameter θ represents the expected value of the random variable M, and models the number of latent risks leading to the relevant event. The Poisson distribution parameters are easily interpreted for the purposes of risk-profile comparison: groups with a larger number of factors leading to recovery are more susceptible to recovery. In this case, we may also say that these are the groups with the highest risk of recovery.

Tables 4 and 5 show the estimated parameters and allow easy comparison of the Poisson parameter estimates across customer groups. Therefore, the results shown in Tables 4 and 5 support the data shown in Tables 1, 2 and 3. We find that the risk profiles of customers with "contracted amount range 2" and "behavior score range 2" show the highest estimated values for θ and, therefore, the most chance of effective implementation of the credit recovery process.

Equipped with the three parameters estimated by the competing risks model, Θ = (θ, γ, β), we show on Tables 6 and 7 the values for the survival of contracts for 12, 18 and 24 month intervals. According to our model's development, $S_Y$ (12months) concerns the probability of a defaulted contract being recovered after 12 months.

Since the contract tracking and collection period is capped at 24 months, the values calculated in $S_Y$ (24months) represent the probability of non-recovery of the non-performing contracts at the end of the 24-month period set for collection efforts. Note that these values, seen in table 6 column $S_Y$ (24months) and in Table 7 as well, are

### Table 4

| Group | Γ | β | θ | Exp(-θ) |
|---|---|---|---|---|
| Value Range I | 1,157 | 18,762 | 0,614 | 0,510 |
| Value Range II | 1,157 | 18,762 | 0,871 | 0,418 |
| BS Range I | 1,260 | 23,152 | 0,413 | 0,661 |
| BS Range II | 1,260 | 23,152 | 1,422 | 0,241 |

### Table 5

| Group | Subgroup | Γ | β | θ | Exp(-θ) |
|---|---|---|---|---|---|
| Value Range I | BS Range I | 1,297 | 28,504 | 0,541 | 0.581 |
| | BS Range II | 1,297 | 28,504 | 1,458 | 0,232 |
| Value Range II | BS Range I | 1,304 | 18,551 | 0,544 | 0,580 |
| | BS Range II | 1,304 | 18,551 | 1,849 | 0,157 |

very close to the non-recovery values initially presented in tables 2 and 3, respectively.

Graphs 1 and 2, next, help compare the recovery risk profiles of the combined profiles formed by "Behavior range" and "contracted amount range".

As expected, the greater number of latent competing risks for the occurrence of credit recovery is related to the greater chance – or risk – of occurrence of the event of interest (recovery).

Finally, Box 1 shows the degree of risk associated with the implementation of the recovery and collection process by increasing order.

### Box 1

| |
|---|
| Low: FX-BS1 combined with FX-CV1 |
| Lower Medium: FX-BS1 combined with FX-CV2 |
| Higher Medium: FX-BS2 combined with FX-CV1 |
| High: FX-BS2 combined with FX-CV2 |

### Table 6

| Group | $S_Y$ (12months) | $S_Y$ (18months) | $S_Y$ (24months) | % Unrecovered |
|---|---|---|---|---|
| Value Range I | 75,89% | 68,56% | 63,65% | 63,19% |
| Value Range II | 67,63% | 58,56% | 53,70% | 53,41% |
| BS Range I | 86,39% | 80,74% | 76,46% | 76,27% |
| BS Range II | 60,46% | 47,93% | 39,74% | 40,39% |

### Table 7

| Group | Subgroup | $S_Y$ (12months) | $S_Y$ (18months) | $S_Y$ (24months) | % Unrecovered |
|---|---|---|---|---|---|
| Value Range I | BS Range I | 86,04% | 79,51% | 74,22% | 74,06% |
| | BS Range II | 66,68% | 53,91% | 44,78% | 45,02% |
| Value Range II | BS Range I | 79,03% | 71,45% | 66,36% | 66,17% |
| | BS Range II | 44,94% | 31,91% | 24,83% | 25,43% |

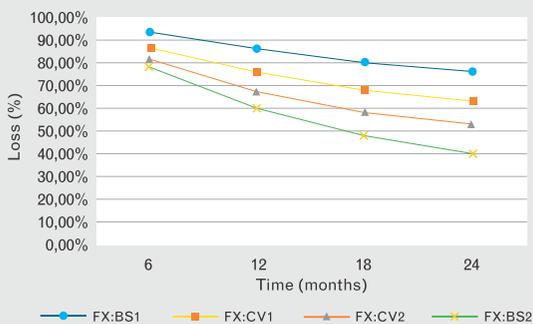

**Graphic 1** Probability of non-recovery

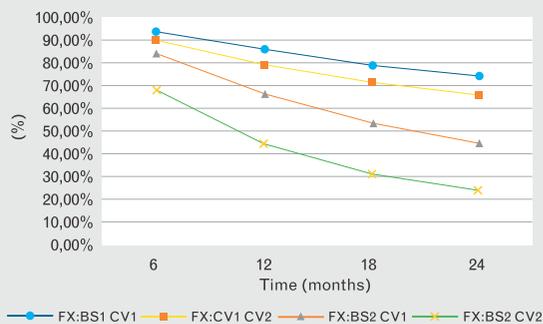

**Graphic 2** Probability of non-recovery

## 4. Conclusion

This article presents a new way to measure the efforts of collection departments, which, generally speaking, are present in every credit-granting industry. We therefore attempt to provide an additional tool to be used jointly with existing methods in use in process of collection and recovery.

The purpose of a great collection policy is to direct where to employ more effort and, on the other hand, where there is no need to do it with excessive expenditure of resources, resulting in a structured policy and economic recovery. We applied the statistical method known as competing latent risks modeling, as intended, and were able to compare groups of customers according to the probability of recovery of their non-performing loans.

We thus expect that, with the combination of a new additional statistical tool applied to the recovery process, credit lenders may pursue the objective of maximizing the collection process, with an immediate reduction of the losses arising from their financing activities.

## 5. Acknowledgments

This study was funded by CNPq and FAPESP, Brazil.

**Authors**

### Mauro Ribeiro de Oliveira Júnior

Has a Bachelor's Degree in Mathematics from Universidade Federal de São Carlos (2003), a Master's Degree in Mathematics from Universidade Estadual de Campinas (2006) and an MBA in Risk Management from FIPECAFI (2012). Is currently a Doctoral Candidate in Statistics at UFSCar. E-mail: *mauroexatas@gmail.com*

### Francisco Louzada

PhD of Statistics from Oxford University (1998), Master of Computer Sciences and Computing Mathematics from Universidade de São Paulo (1991), Bachelor of Statistics from Universidade Federal de São Carlos (1988). Currently a Tenured Professor at Universidade de São Paulo. E-mail: *louzada@icmc.usp.br*